# Evaluation of the Automatic Classifier AutoSense Sky OS for Pediatric Cochlear Implant Users using a Virtual Classroom Environment

Maartje M. E. Hendrikse[a], André Goedegebure[a], Kars R. Tjepkema[a], & Jantien L. Vroegop[a]

[a]Department of Otorhinolaryngology and Head and Neck Surgery, Erasmus MC

University Medical Center Rotterdam, Dr. Molewaterplein 40, 3015 GD Rotterdam, The Netherlands

**All correspondence should be addressed to:**

Maartje M. E. Hendrikse, Erasmus Medical Center, Department of ENT, Room Nt-310, P.O. Box 2040,

3000 CA Rotterdam, The Netherlands. E-mail: m.hendrikse@erasmusmc.nl

# Abstract

This study evaluated whether AutoSense Sky OS, an automatic classifier used in pediatric Advanced Bionics cochlear implants (CIs), improves speech intelligibility from the front without significantly impairing spatial awareness (i.e., sound detection and identification from all directions), with the main focus on the classroom situation. A double-blind crossover study was conducted with 12 pediatric CI users aged 7–16 years. Participants tested two settings: one with AutoSense Sky OS activated and the other with an omnidirectional microphone mode. Speech intelligibility from the front and spatial awareness were assessed in a virtual classroom environment using a Virtual Reality headset and a 12-loudspeaker array. Participants also rated the performance of each setting during a take-home period. A technical evaluation revealed that AutoSense Sky OS activated its directional microphone mode in the virtual classroom, and during the listening test this significantly improved speech intelligibility from the front while significantly reducing spatial awareness. While not all participants reported noticeable differences between settings in daily life, 8 out of 12 children preferred "AutoSense on", perceiving improved speech intelligibility in noisy environments. Conversely, some participants preferred "AutoSense off" due to poorer localization and speech intelligibility from behind with "AutoSense on". In conclusion, the automatic classifier provided improved speech understanding in noisy conditions such as a classroom, at the cost of a slightly reduced spatial awareness. The preference for AutoSense Sky OS among CI children and their parents appears to be a matter of individual trade-offs between improved speech intelligibility in noisy conditions and reduced spatial awareness.

# 1 INTRODUCTION

The ability to comprehend speech in the presence of background noise, a phenomenon known as the cocktail party effect, represents a significant challenge for individuals with hearing impairments (Bronkhorst, 2000). Hearing devices with directional microphones offer an additional advantage over amplification alone, as they can increase the signal-to-noise ratio (SNR) by attenuating sounds from directions other than the frontal direction (Bentler, 2005). Nevertheless, the advantage of directional microphones in everyday life cannot be reliably predicted by standard laboratory tests, given that the benefit of directional microphones is highly dependent upon the characteristics of the listening environment (Cord et al., 2004). Factors such as a large distance between talker and listener, an off-axis position of the talker, small spatial separation between signal and noise sources, and reverberation, reduce the benefit of directional microphones. Furthermore, directional microphones attenuate off-axis targets, which can impair the ability to orient to new sound sources if the attenuation of off-axis signals is too strong (Archer-Boyd et al., 2018). Therefore, the use of directional microphones should be limited to appropriate situations.

Identifying listening situations where the directional microphone could be beneficial and switching manually between programs with omnidirectional or directional microphone mode is challenging for many hearing device users (de Graaff et al., 2018; de Graaff et al., 2024). To address this problem, many modern hearing devices are equipped with an automatic classifier that can switch automatically between programs by analysing the acoustical characteristics of the environment. The efficacy of such automatic classifiers in selecting appropriate settings for varying circumstances remains largely unstudied. However, existing research indicates that there are differences between manufacturers (Husstedt et al., 2017; Yellamsetty et al., 2021). Moreover, automatic classifiers are designed to optimize speech perception, whereas hearing aid users may prefer a program that offers enhanced listening comfort (Übelacker & Tchorz, 2015).

Children encounter different listening situations than adults and therefore have different hearing needs. For example, children may be addressed more frequently from behind when they are engaged in play. In addition, children frequently encounter listening situations in the classroom. Directional microphones may be beneficial in the classroom for frontal instruction or individual work, which accounts for 22-35% of classroom time (Feilner et al., 2016). However, a remote microphone that streams the teacher's voice directly to hearing devices is expected to provide greater benefits in these situations (Wolfe et al., 2020). In contrast, during group work and interactive lessons—which account for 34% of classroom time (Feilner et al., 2016)—multiple, rapidly changing signals of interest arise as different individuals are talking. Noise levels during group work are considerably higher than those observed during frontal instruction and individual work, with reported $L_{A,eq}$ values of 72.9 dB during group work, compared to 61.2 dB during frontal instruction and 64.6 dB for individual work (Shield & Dockrell, 2004). Consequently, enhancing the SNR during group work and interactive lessons is particularly critical. A remote microphone is not appropriate for situations with rapidly changing talkers, and it is therefore unlikely to provide much improvement during group work and interactive lessons. In contrast, when using a directional microphone, the target signal for which the SNR is improved can shift as listeners turn their heads. As a result, in group work and interactive lesson settings, a directional microphone may prove more beneficial than a remote microphone. In a study by Ricketts et al. (2017), trained observers judged a directional microphone mode to be optimal during 33% of the active listening time in a classroom.

The advantages of a directional microphone for speech intelligibility from the front have been researched in children using hearing aids (e.g., Wolfe et al., 2017) or cochlear implants (CIs) (e.g., Ching et al., 2024; Ernst et al., 2019). Wolfe et al. (2017) reported a 21-28 percentage point improvement in frontal sentence recognition in children with hearing aids using adaptive directional microphones, despite an 8-12 percentage point reduction for speech from behind. Wolfe et al. (2017) suggest that because the improvement is larger than the decrement, the benefits of using a directional-microphone mode outweigh the drawbacks. However, this may be a somewhat simplistic representation of reality, and other important factors should also be considered. During group work and interactive lessons, children should be able to switch their attention to a new speaker if necessary. This shift in attention is often accompanied by a corresponding head movement, which allows the directional microphone to point towards the new speaker and thereby improve the SNR of the new speaker. Therefore, when using directional microphones in noisy situations such as group work or interactive lessons, children should still be able to: 1) detect when someone starts talking from a different direction, and 2) get enough information to determine whether it is necessary to switch their attention to the new speaker. The term "spatial awareness" is used to describe the ability to detect and identify sound sources in the surrounding environment.

If the off-axis attenuation of the directional microphone is too strong, spatial awareness may be reduced, which may lead to less interaction with the environment. Although the cost of spatial attention switching has been investigated in children in classroom situations (Loh, Fintor, et al., 2022), the ability of hearing-impaired children to detect an off-axis signal and determine whether it is of interest remains unexplored. Furthermore, the potential negative impact of using a directional microphone on spatial awareness has yet to be evaluated. Additionally, there is a limited understanding of the use of automatic classifiers in children who use hearing aids or cochlear implants. A study by Ganek et al. (2021) evaluated the classification accuracy of an automatic classifier used in CIs from the brand Cochlear (SCAN) in children, but not the transient behavior of the SCAN classifier. The effect of the SCAN classifier on real-world functional performance was also evaluated in CI children (Ching et al., 2024). Wolfe et al. (2017) evaluated the real-world functional performance of an automatic classifier used in Phonak hearing aids and Advanced Bionics CIs (AutoSense) in children who used hearing aids. The two studies identified comparable ratings for the real-world functional performance with and without an automatic classifier, although there was a preference for AutoSense in the study by Wolfe et al. (2017). Since the study by Wolfe et al., a pediatric version of AutoSense (AutoSense Sky OS) has been developed, which has not been evaluated in CI children yet.

In the current study we therefore wanted to evaluate the potential advantages and disadvantages of the automatic classifier AutoSense Sky OS, which we will refer to as "AutoSense", on the auditory function of CI children. First, we aimed to perform a technical evaluation to determine how AutoSense classified several complex acoustic conditions (Experiment 1). Secondly, the objective was to evaluate the impact of AutoSense on speech intelligibility from the front and spatial awareness in a classroom setting with CI children (Experiment 2). Finally, we aimed to evaluate the everyday experience of CI children with AutoSense.

# 2 METHODS

## 2.1 VIRTUAL AUDIOVISUAL CLASSROOM

An audiovisual classroom situation was implemented in Virtual Reality (VR) in order to facilitate measurements in a realistic setting and to make the test more engaging for children. Furthermore, with the use of VR, high ecological validity could be obtained within a controlled laboratory environment.

The virtual acoustic environment of the classroom was based on recordings of babble noise and room impulse responses made in a real classroom. Babble noise recordings were made with a first-order ambisonics microphone (Zoom H3-VR) while children aged 10-11 were engaged in group assignments. From the babble noise recordings, a 124-second segment was selected that exhibited a comparable sound level and minimal intelligible speech content. This selected fragment was used as diffuse noise in the virtual classroom. As evidenced in the literature on the acoustic properties of classroom environments, sound levels are dependent upon the activity. As reported by Shield & Dockrell (2004), the average $L_{A,eq}$ values were found to be 56.3 dB during silent reading, 61.2 dB during frontal instruction, 64.6 dB during individual work (72.2 dB with children walking around) and 72.9 dB during group work (76.8 dB with children walking around). Long-term $L_{A,eq}$ values in classrooms and preschools are around 66 dB (Loh, Yadav, et al., 2022). For the virtual classroom, we chose an average sound level of 65 $dB_A$ for the diffuse babble noise, which is similar to the long-term $L_{A,eq}$ values reported by Loh et al. (2022). Additionally, point noise sources were introduced, including the opening of a pencil case, clicking with a pen, dropping a pencil, and flipping pages.

The layout and dimensions of the actual classroom were used to implement first-order reflections on the walls, floor, ceiling, and nearby tabletops. The room impulse response recordings were employed to optimize the parameters of a feedback-delay network for the late reverberation, with a T60 of 0.43 s. The ANSI Standard S12.60 (ANSI & ASA, 2010) recommends a maximum reverberation time of 0.6 s in an unoccupied, furnished classroom with a volume less than 10,000 cubic feet (283 $m^3$). The actual classroom was found to comply with this standard.

For the listening test in Experiment 2, digit triplets were presented from the front, at changing sound levels, and stimuli of animal names (Dutch words for tiger and rabbit) were presented at 60 $dB_A$ and 65 $dB_A$ from 0°, ± 90°, ± 120° and 180°.

A matching visual environment was created in Unreal Engine (version 4.27, Epic Games), see Figure 5. Virtual characters uttering the speech stimuli were animated with mouth movements during speech. However, the animations were always the same, ensuring no audiovisual intelligibility benefit. The VR controllers were visible in the visual environment and featured an attached animal icon to identify which controller's button to press.

The room impulse response recordings made in the real classroom, the first-order ambisonics babble noise recording and the 3D environment of the room are published (Hendrikse & Tjepkema, 2025).

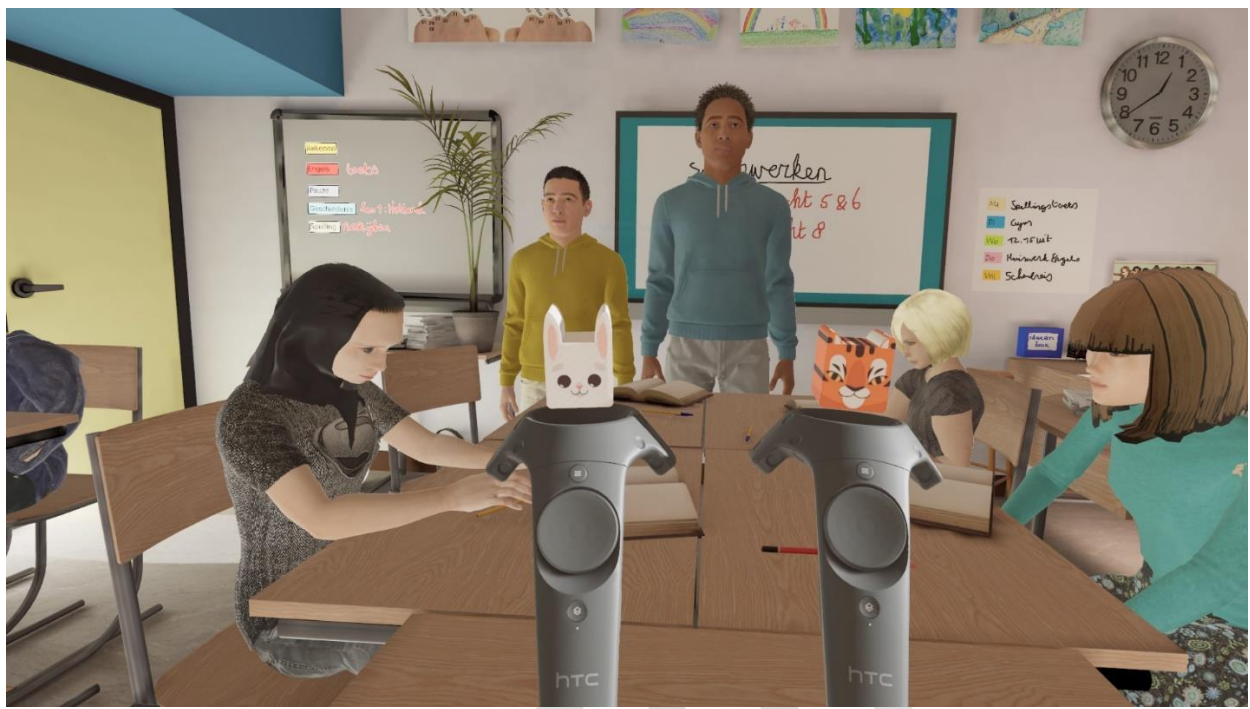

*Figure 1: Screenshot of the VR classroom seen by the participants when wearing the VR headset. The animated character in the front with the blue sweater spoke the digit triplets, presented from the loudspeaker at 0°. The animated character in the front with the yellow sweater spoke the animal stimuli, also presented from the loudspeaker at 0°. This same character with different colored sweater was also located at ± 90°, ± 120° and 180°, and the animal stimuli were also presented from the loudspeakers at these angles. The animal figures attached to the controllers indicated which controller's button to press when hearing "tijger" (tiger) or "konijn" (rabbit).*

## 2.2 EXPERIMENT 1 "TECHNICAL EVALUATION OF AUTOMATIC CLASSIFIER"

In order to evaluate AutoSense, we utilized the method described by Husstedt et al. (2017) to identify the activated categories of the automatic classifier in the virtual acoustic classroom situation. The technical evaluation of AutoSense was also performed in two different speech-in-noise situations in a virtual acoustic pub, and in a virtual acoustic traffic situation on a street intersection. See Appendix A for a detailed description and the results of the pub and street situations.

The method described by Husstedt et al. (2017) necessitates the acquisition of multiple recordings of a hearing device's acoustical output under identical acoustic conditions. This is done in order to assess the impact of a modification in the amplification settings within one of the automatic classifier categories on the acoustic output. Since this is only feasible with hearing aids and not with CIs, the technical evaluation was conducted with Phonak hearing aids that had the same AutoSense classifier as Advanced Bionics Sky CIs.

AutoSense employs a classification system comprising the following categories: "calm situation", "speech in noise" (with subcategories "speech in loud noise" and "speech in car"), "comfort in noise", "comfort in echo", and "music". For further details on AutoSense, please refer to Appendix B. We expected AutoSense to select the "speech in noise" category in the virtual acoustic classroom situation, which activates the directional microphone mode.

### 2.2.1    Hearing aid configuration

The technical evaluation was conducted using a pair of Phonak Sky Link M hearing aids, which were placed on both ears of a dummy head. The Sky hearing aids were equipped with AutoSense Sky OS version 3.0. The hearing aids were programmed for a fictitious 11-year-old patient with a symmetric moderate/severe steep sloping hearing loss consistent with the S3 standard audiogram from Bisgaard, Vlaming, & Dahlquist (2010), using junior mode for that age and DSL v5a pediatric fitting rule (Scollie et al., 2005). A universal tip was used to place the hearing aids on the dummy head. A feedback & real ear test was performed with the hearing aids placed on the dummy head, to reduce feedback.

### 2.2.2    Measurements and Analysis

The acoustic output of the hearing aids placed on the dummy head was recorded with real-ear measurement equipment (Affinity Compact, Interacoustics), by inserting the probe tube in the ear canal. The dummy head was a custom-made polystyrene model with 3D-printed ears and ear canals. The interaural distance of the dummy head (17.5 cm) was comparable to that of a real head. In the method proposed by Husstedt et al. (2017), the sole aspect of interest is the difference between two recordings. Therefore, the discrepancy between the dummy head and a real head with respect to the acoustic properties of the ear canal is inconsequential. The recording was conducted on a single ear at a time, with the right ear selected for analysis.

In accordance with the methodology proposed by Husstedt et al. (2017), a reference recording of the acoustic output of the hearing aids was obtained in the virtual acoustic classroom situation. Subsequently, the overall gain in one of the classifier categories was reduced by 15 dB as a marker, after which the recording of the acoustic output of the hearing aids was repeated. This procedure was carried out for all classifier categories. In the virtual acoustic classroom situation, the digit triplets and animal stimuli from the listening test described in Experiment 2 (section 2.3.3) were played at the time points and sound levels that were used in a test with a pilot participant.

A discrepancy between the reference recording and a recording with a marker indicated that the gain reduction affected the hearing aid output, thereby suggesting that the classifier category with the marker was likely active. To analyze the differences between recordings, the recorded signals were filtered with an eight-order 1/3-octave filter bank with a passband ranging from 100 Hz to 8 kHz. The filtered signals were divided into analysis windows of one second in duration, thus enabling the examination of alterations in active classifier categories over time in the context of the dynamic acoustic conditions. The differences in decibels between the reference recording and recording with a marker were calculated per octave band and per analysis window. As outcome measure, the mean difference in dB with the reference recording was calculated over all octave bands per analysis window for the recordings with a marker in each of the classifier categories. Mean differences in dB where the sound level of the recording with the marker was less than 1 dB lower than the reference recording were disregarded, because this could be attributed to noise in the recordings.

## 2.3   Experiment 2 "Crossover Intervention Study"

A crossover intervention study was performed to investigate the effect of AutoSense in everyday performance of pediatric CI users, with focus on the classroom situation.  Since AutoSense switches to directional microphone mode in speech in noise situations, we expected that speech intelligibility from the front would improve, whereas spatial awareness might be negatively affected, although not severely. This specific aspect was investigated by using a novel listening test that we implemented in the VR classroom situation. Additionally, children and parents provided feedback on their experience with AutoSense on and off during a take-home period via questionnaires.

### 2.3.1   Participants

A sample size calculation indicated that a total of 12 participants were required. Therefore, a total of 12 hearing-impaired children between the ages of 7 and 16 years (mean age: 12 years), who were wearing one or two CIs, were included in the study. The study population consisted exclusively of children with Advanced Bionics Sky CI M90 sound processors. All participants had been using their CIs for more than six months at the start of the study and had phoneme perception scores in quiet exceeding 70% at 65 dB SPL. Exclusion criteria were additional disabilities that could be expected to interfere with VR evaluation, such as vision problems or epilepsy.

Written informed consent was obtained from both parents, and from children aged 12 years and above. For children under the age of 12 years, only their parents signed the informed consent form. Travel expenses incurred for attendance at study appointments were reimbursed, and the children received a gift voucher as a token of appreciation for their participation. The study protocol was approved by the Erasmus Medical Center Ethics Committee (reference number NL85237.078.23, 16-06-2023).

### 2.3.2   Conditions

Two settings of the CIs were compared in this study. The first setting, "AutoSense off", was equivalent to the "calm situation" category of AutoSense, which means that the CI microphones were always in omnidirectional mode or so-called "real ear sound". In the second setting, "AutoSense on", AutoSense was active and changed the noise reduction parameters and microphone mode according to the category that was selected based on the acoustic conditions (see Appendix B). The order of the settings was blinded for both the participants and the experimenter responsible for conducting the listening test.

### 2.3.3   Listening Test

The listening test consisted of two tasks. The children were asked to repeat digit triplets spoken by a virtual teacher positioned in front of the virtual classroom, in order to assess speech intelligibility from the front direction. Secondly, the children were instructed to press the button on the VR controller that corresponded to the animal figure they heard, either "tijger" (tiger) or "konijn" (rabbit), in order to evaluate their spatial awareness. Trials of the two tasks were randomly interleaved, with the restriction that no more than four trials from the same task could be presented consecutively. The time between the trials was randomly assigned, ranging from 3 to 20 seconds. The experimenter pressed a button to make confetti appear in the virtual environment when the children repeated the digit triplets, correctly or incorrectly, as positive reinforcement for the children's cooperation with the test.

### 2.3.3.1 *Speech intelligibility from the front*

To evaluate speech intelligibility from the front, the 50% speech reception threshold (SRT50) was measured using digit triplets from the Dutch digits-in-noise (DIN) test (Smits et al., 2013). The children were asked to repeat all three digits after the presentation of a triplet. The noise level was kept constant at 65 dB$_A$, and the level of the digit triplets was adjusted based on the correctness of the response using a standard one-up-one-down procedure with a step size of 2 dB. The first triplet was repeated at higher sound levels until all three digits were identified correctly, before presenting the second triplet.

The DIN test is used clinically in the Netherlands. It has only limited effects of top-down processes such as linguistic skills and has been shown to provide reliable results in children with hearing loss and CIs (Vroegop et al., 2021). The speech material used in the DIN test consists of a set of 120 unique digit triplet combinations constructed from the digits 0 to 9 uttered by a male speaker, separated by silent intervals and organized into lists of 24 digit triplets. One list of 24 digit triplets was used to measure the SRT50. The SRT50 was calculated by taking the average SNR of trials 5 through 25, where the 25th SNR is the value resulting from the adaptation rule based on the response to the 24th trial. The clinically used DIN test has a masking noise that matches the long-term average speech spectrum of the digit triplets, which is presented from the same loudspeaker as the digit triplets. In this study, diffuse babble noise recorded in a classroom was used. This did not match the long-term average speech spectrum of the digit triplets and was presented over 12 loudspeakers. Accordingly, it was anticipated that the SRT50 values would differ from those clinically measured.

### 2.3.3.2 *Spatial Awareness*

To evaluate spatial awareness, the children were asked to press the button on the VR controller with the corresponding animal figure when they heard "tijger" (tiger) or "konijn" (rabbit). This task was intended to mimic the situation where one of their classmates wants to catch their attention, for instance to ask a question, which is an example of spatial awareness. In such a situation, the children have to detect that a classmate is calling, and have to roughly identify what they are saying, which is mimicked by letting them identify the difference between the two words "tijger" and "konijn". The position of the VR controllers was tracked, and they were visualized in the VR environment. An animal figure was placed on top of the VR controllers in the VR environment to indicate which VR controller corresponded to which animal (see Figure 1). The animal stimuli were presented from 0°, ±90°, ±120° and 180°, at SNRs of -5 and 0 dB$_A$ (60 and 65 dB$_A$). Each animal stimulus was presented once from each location and at each SNR, resulting in 24 trials in total. The SNRs were chosen based on the results from pilot tests, in a range where we did not expect to observe floor or ceiling effects. The animal stimuli, "tijger" and "konijn", were generated using a text-to-speech algorithm, mimicking a child's voice. The children's button presses were recorded, as well as their response times. The children's head movements during the listening test were also recorded. The percentage of trials in which the children pressed any button (either the correct animal or an incorrect one) was calculated as a detection score. Moreover, the proportion of trials in which the children selected the correct animal button was calculated as an identification score, and the reaction time was determined as a measure of their listening effort. Only responses received within four seconds of the presentation of the animal stimulus were considered valid.

*2.3.3.3 Setup*

For sound presentation, the setup (Figure 2) consisted of a loudspeaker ring with 12 loudspeakers (Genelec 8020), controlled by a Linux PC (DELL Precision 3640) running an acoustic simulation program (TASCAR 0.229 (Grimm et al., 2019), open-source software) via a sound card (Motu 24Ao). The sound card was connected to a PCIe card (RME HDSPe RayDat) with ADAT cables. The equipment was calibrated with a decibel meter (Brüel & Kjær type 2250), to ensure that the sound presentation was at the intended sound levels.

For presentation of the visual stimuli, a VR headset (HTC Vive Pro Eye) was used, controlled by a Windows PC (HP Z1 G8) running Unreal Engine 4.27 (Epic Games Inc., 2021) with the virtual classroom (Figure 1).

The listening test was programmed in MATLAB (The MathWorks Inc., 2021) on the Linux PC, which sent Open Sound Control (OSC) messages locally or via the Wi-Fi network to TASCAR and Unreal Engine to start simultaneous playback of audio and visual stimuli. Unreal Engine recorded the location and rotation of the VR headset and button presses on the VR controllers. These data were sent as OSC messages to TASCAR. TASCAR logged the presentation times of the stimuli and the button presses and head movements on the same timeline, which was saved as a MATLAB file at the end of each test.

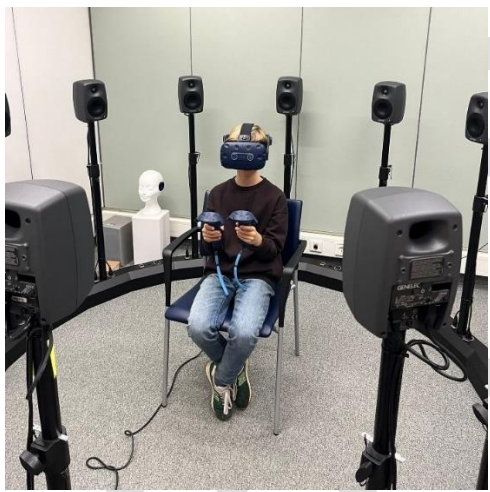

*Figure 2: Picture of setup with loudspeaker-ring and child wearing the VR headset.*

### 2.3.4 Take-home Period and Questionnaires

During the take-home period, the children and their parents tried both the "AutoSense on" and "AutoSense off" setting of the CIs, for two weeks each. After two weeks, a remote fitting session was planned to change to the other setting. During the take-home period, the children and parents were asked to describe their experience with the two settings in a structured journal and fill in a questionnaire at the end of every two weeks.

The children and their parents were asked to keep a journal where they described their experience with the settings in a number of specific situations, as was done by Vroegop et al. (2017). They were asked to do this for a 1-on-1 conversation with and without background noise, listening to speech from a distance (i.e., watching TV or at school), a group conversation with and without background noise, and for a traffic situation. For each category, they selected one specific situation they encountered. In the journal, they were asked to write a description of the situation, and to rate the sound quality, their general experience, and speech intelligibility or sound localization in the traffic situation. In the first two weeks, they rated the sound quality, speech intelligibility and localization by giving an absolute score on an 11-point scale ranging from very bad to very good. In the second two weeks, they gave a relative rating on the 11-point scale to indicate whether the setting they tried second was better or worse than the first.

The questionnaire the children and their parents were asked to complete was a selection of questions from the pediatric version of the Speech, Spatial, and Qualities of Hearing Scale (SSQ). The pediatric version of the SSQ (Galvin & Noble, 2013) was translated to Dutch and validated by Betthyany et al. (2023). A number of questions were selected that were deemed important for this study because they describe speech-in-noise or spatial awareness situations: Speech 1, 4, 6, 7, and 8; Spatial 1, 3, 4, and 6; Quality 2 and 7. The children and parents were asked to complete the questionnaire twice: once for their experience during the first two weeks, and again for the second two weeks. During the second two weeks, they were allowed to use their answers from the first two weeks as a reference.

### 2.3.5 Randomization and Test Sessions
Participants performed the listening test with both test conditions ("AutoSense on" and "AutoSense off") twice, both at the beginning and at the end of the take-home period, so that a test-retest comparison could be done. The duration of the listening test was approximately eight to ten minutes for one condition, and there was a short break between the listening test with the first and second condition. To ensure the integrity of the study, the order of the settings during the listening tests and take-home period was counterbalanced across participants through a Latin square design. With a sample size of twelve, each possible order was administered on two occasions.

### 2.3.6 Analyses
The effect of AutoSense condition (off or on) and test session on the outcome measures was analyzed. The outcome measures for the listening test were the SRT50 value for the DIN test and the percentage of detected and identified animal stimuli, as well as the reaction time for the animal stimuli. The data were analyzed using generalized linear mixed-effects models (GLMMs), which are capable of handling non-normally distributed variables and missing data points. GLMMs were fitted in R (version 4.3.2; R_Core_Team, 2023) using the *lme4* package (version 1.1.35.5; Bates et al., 2015) with Maximum Likelihood and Nelder-Mead optimizer. The resulting models were then checked for goodness of fit, collinearity, and overdispersion using the *performance* package (version 0.10.5; Lüdecke et al., 2021) and described using the *report* package (version 0.5.9; Makowski et al., 2020). Standardized parameters were obtained by fitting the models on a standardized version of the dataset, so that the effect sizes of the parameters could be compared. 95% Confidence Intervals (Conf. Int.) and p-values were computed using a Wald t- or z-distribution approximation to determine which parameters had a significant effect.

With regard to the questionnaires, the outcome measures are the mean SSQ scores for both settings in the domains of speech, spatial hearing, and sound quality. From the journal, the scores for the direct comparison between both settings in different listening situations are taken as outcome measures.

# 3   RESULTS

## 3.1   EXPERIMENT 1 "TECHNICAL EVALUATION OF AUTOMATIC CLASSIFIER"

Figure 3 shows the acoustic conditions in the virtual acoustic classroom situation and the mean difference in dB with the reference recording for each of the classifier categories of AutoSense. The results for the virtual acoustic pub and street situations can be found in Appendix A.

The results show that sometimes two categories were active at the same time. AutoSense did not switch instantaneously between categories, as this would induce artefacts, but made a weighted mix between the settings of the two categories for some time when switching. This applies to both the gain settings and the microphone mode.

As can be seen in Figure 3, the "speech in noise" category was active in the classroom situation, as expected. There was an initialization period of about 17 seconds in which AutoSense switched from the default start-up category "calm situation" to "speech in noise". Unexpectedly, AutoSense sometimes switched to "comfort in noise". This was always at the same time point in the diffuse babble noise recording, which was 124 seconds long and was looped. At this time point in the babble noise recording one of the children in the classroom laughed loudly.

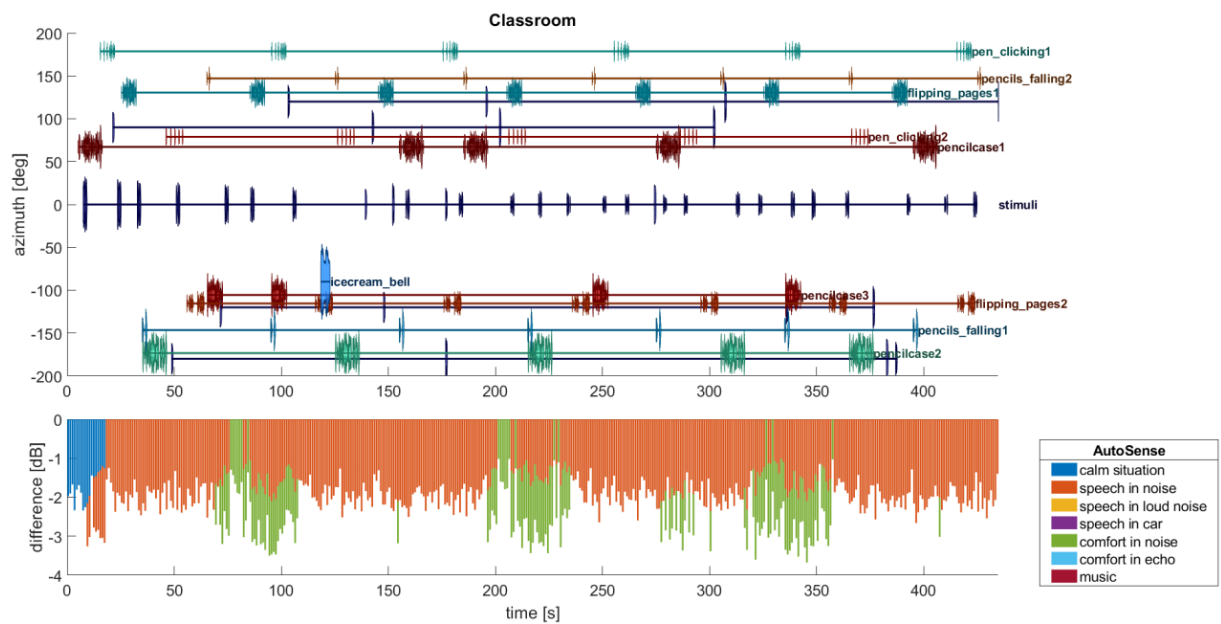

*Figure 3: Results of the technical analysis of the automatic classifier in the classroom. The top graph describes the acoustic conditions and shows the location in azimuth relative to the listener and the sound level (1 dB/degree) at the location of the listener. The bottom graph shows the difference in dB between the reference recording and the recordings with a marker in each of the classifier categories for Sky hearing aids with AutoSense Sky OS 3.0.*

## 3.2   Experiment 2 "Crossover Intervention Study"

### 3.2.1   Listening Test

One participant opted to withdraw from the study during the take-home period, stating a dislike for the "AutoSense on" setting. Consequently, only the participant's listening test results from the initial session, along with their final choice and motivation, were included in the analyses.

In the subsequent analyses, only the percentage of animal stimuli that was correctly identified was used as an outcome measure. We chose not to analyze the percentage of animal stimuli that was detected, because only few confusions were made by the children in their animal-name responses ("tijger" instead of "konijn" and vice versa), on average 0.43 confusions out of 24 stimuli, at most 3 confusions out of 24 stimuli in one test. Thus, there was not much difference between the percentage of animal stimuli that was detected and the percentage of animal stimuli that was correctly identified.

### 3.2.1.1 Speech intelligibility from the Front

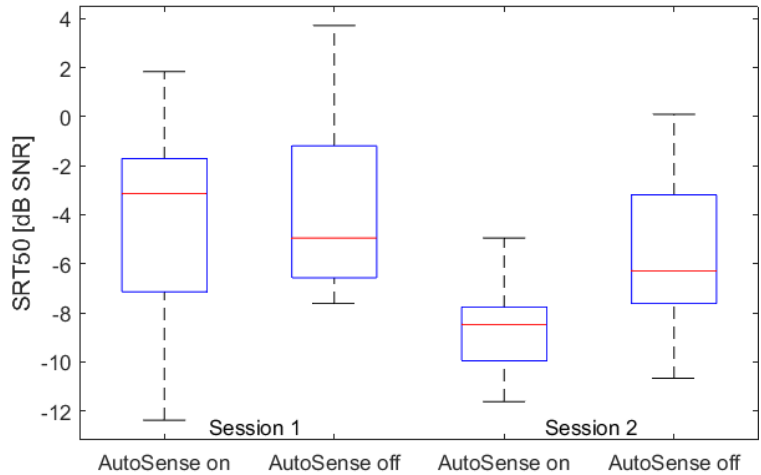

*Figure 4: SRT50 values for the DIN test with AutoSense on/off and for the two test sessions.*

Figure 4 shows the SRT50 values for the DIN test with AutoSense on/off and for the two test sessions. When a linear model with a Gaussian distribution was fitted, the resulting model was overdispersed. This was likely due to the data not approximating a normal distribution closely enough, as a consequence of the relatively small sample size. The data distributions in Figure 4 exhibit a slight skewness, so we fitted a general linear mixed model (GLMM) of the Gamma family with an inverse link function to predict SRT50 (shifted to positive values) with session, condition, and their interaction. The model included a random intercept for each participant. The model showed weak explanatory power, with a conditional $R^2$ of 0.03 and a marginal $R^2$ of 0.02, indicating that the fixed effects explained only a small portion of the variance. No overdispersion and no collinearity were observed after model adjustment. The results for the fixed effects, including the interaction, are presented in Table 1, with corresponding 95% confidence intervals (Conf. Int.) and p-values.

*Table 1: GLMM for SRT50 values.*

| Parameter | Std. beta | 95% Conf. Int. | Wald t | p |
|---|---|---|---|---|
| Session | 0.02 | [-0.02, 0.05] | t(40) = 1.00 | 0.326 |
| Condition | 2.79e-3 | [-0.03, 0.03] | t(40) = 0.20 | 0.846 |
| Session * Condition | 0.09 | [0.03, 0.15] | t(40) = 3.04 | 0.004* |

The GLMM shows a significant interaction effect of session and condition. As can be seen in Figure 4, the SRT50 with AutoSense on was significantly lower (better) than with AutoSense off, but only in the second test session. The mean difference in SRT50 between "AutoSense on" and "AutoSense off" in the second test session was 3.3 dB SNR.

### 3.2.1.2  Percentage of correctly identified animal stimuli

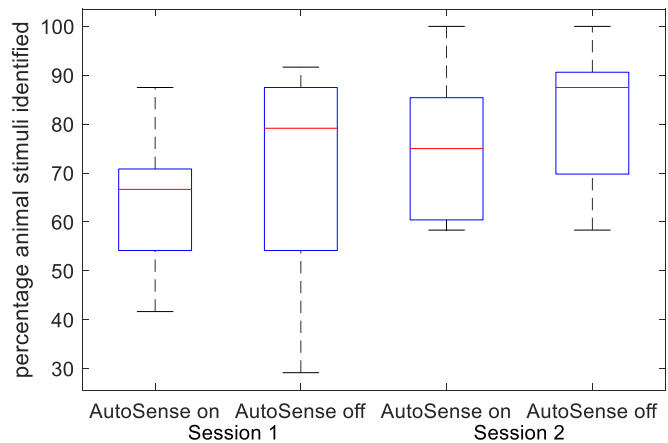

*Figure 5: Percentage of correctly identified animal stimuli with AutoSense on/off and for the two test sessions.*

Figure 5 shows the percentage of animal stimuli that was correctly identified by the participants with AutoSense on/off and for the two test sessions. A GLMM with binomial distribution and logit link function was fitted to predict the answer to each trial (whether the stimulus was correctly identified or not). The model included session, condition, stimulus angle, stimulus SNR and stimulus animal as fixed effects and included a random intercept for each participant. The model showed moderate explanatory power, with a conditional $R^2$ of 0.21 and a marginal $R^2$ of 0.13, indicating that the fixed effects explained a moderate portion of the variance. No overdispersion and no collinearity were observed.  The results for the fixed effects are presented in Table 2, with corresponding 95% confidence intervals (Conf. Int.) and p-values.

*Table 2: GLMM for identification of animal stimulus, per trial.*

| Parameter | Comparison | Std. beta | 95% Conf. Int. | Wald z | P |
|---|---|---|---|---|---|
| Session | 1 vs 2 | 0.62 | [0.33, 0.91] | 4.18 | < .001* |
| Condition | AutoSense off vs on | -0.40 | [-0.68, -0.11] | -2.73 | 0.006* |
| Angle | 0° vs 90° | 0.64 | [0.15, 1.12] | 2.56 | 0.010* |
| | 0° vs 120° | 0.54 | [0.06, 1.02] | 2.19 | 0.029* |
| | 0° vs 180° | 0.14 | [-0.32, 0.60] | 0.59 | 0.554 |
| | 0° vs 240° | 0.41 | [-0.07, 0.88] | 1.69 | 0.092 |
| | 0° vs 270° | 0.67 | [0.18, 1.16] | 2.69 | 0.007* |
| SNR | -5 dB vs 0 dB | 0.81 | [0.52, 1.09] | 5.50 | < .001* |
| Animal | Konijn vs tijger | 0.46 | [0.17, 0.74] | 3.16 | 0.002* |

The GLMM shows significant effects of test session, condition, stimulus angle, stimulus SNR and stimulus animal. Participants identified significantly more animal stimuli correctly in the second test session. Significantly more animal stimuli were correctly identified when AutoSense was off, with a mean difference of 6.4% between "AutoSense off" and "AutoSense on" in the second test session. This was a small effect, and the effects of test session, angle, stimulus SNR, and stimulus animal were all larger, judging from the standardized coefficients in the GLMM (Table 2). The animal stimuli with the higher SNR and with the animal "tijger" were easier to identify. To further investigate the effect of the stimulus angle, a polar plot was made of the percentage of animal stimuli correctly identified per angle (Figure 6). Because of the significant effect of the test session on this outcome measure, only the data from the second test session are included in this plot. The polar plots show that less animal stimuli from 180° were correctly identified when AutoSense was on, on average 29.5% less. A Wilcoxon signed rank test confirmed that this difference between conditions at 180° was significant (p=0.0039).

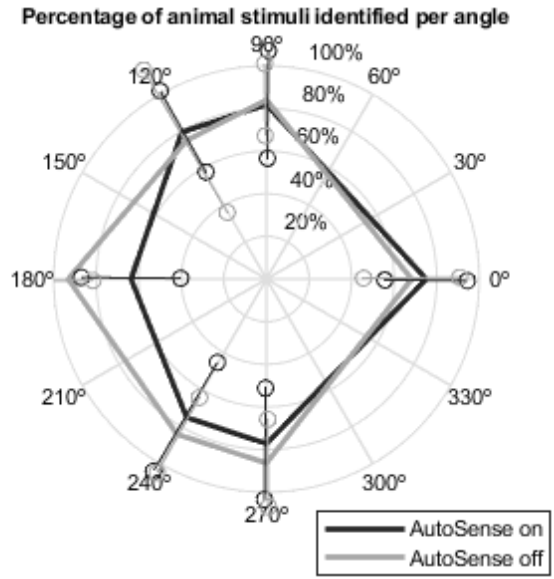

*Figure 6: Average percentage of animal stimuli identified per presentation angle of the stimuli for the conditions AutoSense on (black) and off (grey). Error bars indicate the standard deviation over the participants, only data from the second test session are included.*

### 3.2.1.3    Reaction time to animal stimuli

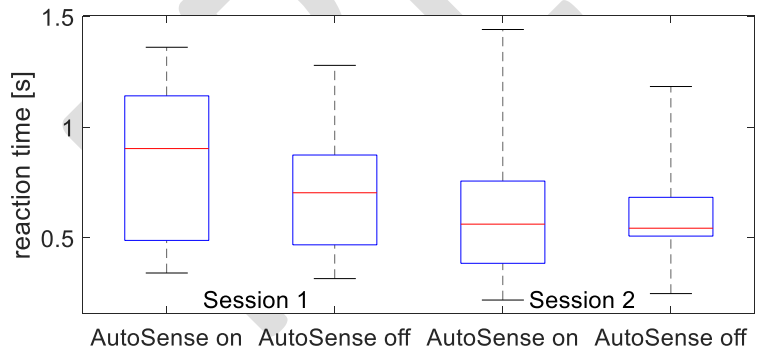

*Figure 7: Reaction time to animal stimuli with AutoSense on/off and for the two test sessions.*

Figure 7 shows the reaction time to animal stimuli with AutoSense on/off and for the two test sessions. A GLMM of the Gamma family with an inverse link function was fitted to predict the reaction time to the animal stimuli per trial with session, condition, and stimulus SNR. The model included a random intercept per participant. The model showed moderate explanatory power, with a conditional $R^2$ of 0.23 and a marginal $R^2$ of 0.03, indicating that the fixed effects explained a small portion of the variance. No overdispersion and no collinearity were observed. The results for the fixed effects are presented in Table 3, with corresponding 95% confidence intervals (Conf. Int.) and p-values.

*Table 3: GLMM for reaction time to animal stimuli, per trial.*

| Parameter | Std. beta | 95% Conf. Int. | Wald t | p |
|-----------|-----------|----------------|--------|---|
| Session | 0.25 | [0.08, 0.41] | t(820) = 2.95 | 0.003* |
| Condition | -0.12 | [-0.27, 0.04] | t(820) = -1.49 | 0.136 |
| SNR | 0.19 | [0.03, 0.34] | t(820) = 2.41 | 0.016* |

The GLMM shows a significant effect of test session and stimulus SNR. Participants were significantly faster to respond to the animal stimuli in the second test session. Moreover, participants were significantly faster to respond to the animal stimuli presented at a higher SNR. However, no significant difference was found for the factor Condition (Autosense on or off).

### 3.2.2    Questionnaires

Five participants did not (completely) fill in the questionnaires, resulting in a significant amount of missing data. Hence, a statistical analysis was not feasible. Instead, we reported observed trends in the questionnaires.

Only 3 out of the 9 participants who completed the SSQ rated the two programs differently for the spatial hearing and quality of hearing questions. Regarding the spatial hearing questions, 2 out of 3 participants that noticed differences preferred "AutoSense on". For the quality of hearing questions, one participant rated "AutoSense on" better, one participant rated "AutoSense off" better, and one participant rated "AutoSense on" better on one question but "AutoSense off" better on the other question. For the SSQ questions related to speech, 6 out of 9 participants noticed differences. Among these six participants, three rated "AutoSense on" better and three rated "AutoSense off" better, although the differences in scores between programs when "AutoSense on" was preferred were larger.

In the journal, the participants had to rate per situation in a direct comparison whether "AutoSense on" or "AutoSense off" was better. The ratings of the seven participants that completed the journal are plotted in Figure 8. On average, the participants noticed little difference between the two programs, with medians and means close to zero. The means indicate a slight preference towards the "AutoSense on" program, and the medians indicate that the majority of the participants preferred "AutoSense on" in "group conversation in noise" situations and "AutoSense off" in traffic situations.

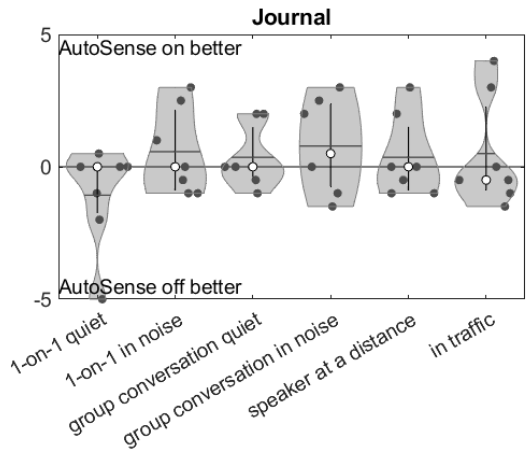

*Figure 8: Direct comparison between programs as rated by participants in the journal. The programs were rated on a scale ranging between -5 (the second program is much worse than the first program) and 5 (the second program is much better than the first program). Scores were corrected for the order of programs during the take-home period, so that a score of 5 corresponds to "AutoSense on is much better". The violin plots show all data points, the 5th and 95th percentiles (vertical black line), the mean (black horizontal line in the violin), and the median (white dot).*

### 3.2.3    Final Choice and Motivation

Eight of the 12 participants elected to keep the setting with "AutoSense on" following the study. Three participants chose to keep the setting with "AutoSense off", and one participant chose the setting with "AutoSense on" but wanted an additional manual program with "AutoSense off" (calm situation program) for use during sports.

With "AutoSense on", participants had an improved perceived and/or measured speech intelligibility in noisy situations (6 participants), a more calm/quiet sound perception (1 participant) and increased alertness in traffic situations (1 participant). However, 3 participants also noticed a poorer localization and poorer ability to understand speech from behind with "AutoSense on". Two participants observed an increase in audibility with the "AutoSense on" setting, but also perceived it as somewhat intense, and experienced headaches. Four out of the 12 participants noticed little or no difference between the two settings.

## 4    DISCUSSION

### 4.1    EXPERIMENT 1 "TECHNICAL EVALUATION OF AUTOMATIC CLASSIFIER"

In general, AutoSense selected a category that seemed appropriate for the acoustic condition that was tested, with a few exceptions. In the virtual acoustic classroom situation, AutoSense sometimes switched to "comfort in noise". Switching to "comfort in noise" activated the omnidirectional microphone mode, with a blend between microphone modes (semi-directional microphone mode) during the switch. Performance differences between the two programs with AutoSense activated and deactivated during these time periods were therefore less pronounced when conducting the listening test in the intervention study. However, this could also happen in real life, and the time periods where AutoSense was fully switched to "comfort in noise" were not that long, so we decided not to change anything in the acoustic conditions for the intervention study. In the street situation, AutoSense switched to "speech in noise" when speech was detected and activated the directional microphone, which could be detrimental for hearing approaching traffic (see Appendix A). In the pub situations, we expected AutoSense to switch to "speech in loud noise", but with the default settings of AutoSense this did not happen (see Appendix A).

Although we followed the methodology described by Husstedt et al. (2017), we modified it slightly to enable us to also examine the transient behavior of the automatic classifier, as opposed to solely focusing on the final result of the classification, as Husstedt et al. did. A noteworthy observation regarding the transient behavior was that AutoSense required approximately 17 seconds to transition to the "speech in noise" program in a classroom setting. In light of these findings, it is recommended that a 17-second interval be observed after the activation of the noise before the presentation of the initial stimulus during a speech-in-noise test. This approach assumes that the babble noise is sufficiently realistic to be recognized as speech by AutoSense, otherwise AutoSense would transition to "comfort in noise" instead of "speech in noise". The 17-second waiting period was incorporated into the listening test in Experiment 2.

## 4.2   EXPERIMENT 2 "CROSSOVER INTERVENTION STUDY"

The observed effects followed the expected pattern: better speech understanding from the front at the cost of poorer spatial awareness with "AutoSense on".

The participants had a significantly lower (better) SRT50 in the listening test in the virtual classroom with "AutoSense on", but this effect was only observed in the second test session. Since there was no significant difference in the SRT50 with "AutoSense off" between the first and second test session, it can be inferred that this improvement was not a training effect associated with the test method. The head movements made by the participants were similar in the first and second test session, so this cannot explain the difference either. Between the test sessions, participants had time to trial the "AutoSense on" setting during the take-home period. Most participants had not used the "AutoSense on" setting prior to the study. Therefore, we believe participants needed an adjustment period to become accustomed to the "AutoSense on" setting.

The superior SRT50 with the "AutoSense on" setting was hypothesized, because the directional microphone that was activated by AutoSense in the virtual classroom improves the SNR for the frontal direction. An analysis of the head movements made by the participants revealed that they were primarily oriented towards the direction of the virtual teacher (positioned at 0°), thereby confirming the SNR benefit for the digit triplets uttered by the virtual teacher. The improved speech intelligibility from the front with the "AutoSense on" setting is in accordance with the findings of Wolfe et al. (2017), who observed a speech intelligibility benefit in children with hearing aids equipped with an older version of AutoSense. A prior study by Ernst et al. (2019) observed an SRT50 improvement of 1.6-4.3 dB in a sentence recognition in noise test with CI children using the same directional microphone algorithm as used in AutoSense, depending on the noise configuration (speech-shaped noise presented from [180°, ±120°, ±60°] or [180°, ±60°, ±30°]) and unilateral/bimodal/bilateral CI use. The SRT50 improvement observed in the current study was of a comparable magnitude (3.3 dB).

The "AutoSense on" setting had a slight but consistent detrimental impact on the spatial awareness in the listening test conducted in the virtual classroom. The percentage of animal stimuli that was identified correctly was significantly lower with "AutoSense on", particularly for the stimuli presented from 180°. While the mean difference between the "AutoSense on" and "AutoSense off" conditions was relatively minor (6.4%) across all angles, the mean difference at 180° was notably pronounced (29.5%). This can be explained by the pronounced notch in the directional beam pattern of the AutoSense microphone system for sounds from the back. A significant effect of test session was observed for both the percentage of correctly identified animal stimuli and the reaction time in the spatial awareness task. Therefore, it appears that a learning effect may be present with regard to the spatial awareness task. As the order of testing the AutoSense conditions was counterbalanced across participants, this learning effect did not impact the outcome of the study.

Unfortunately, the response rate to the questionnaires was low. In the interest of improving the quality of future studies, it is recommended that researchers explore alternative methods for enhancing response rates. One potential approach is the use of mobile applications and ecological momentary assessment (EMA) techniques. The participants who completed the questionnaires in the current study provided similar ratings for the two settings, although some indicated a preference for either "AutoSense on" or "AutoSense off." A comparable outcome was observed with regard to the final selections made by the participants and the rationale behind these choices. Not everyone noticed a difference between the settings, but the majority of the participants chose the "AutoSense on" setting. Some participants preferred AutoSense only in certain situations. There were also participants who noticed adverse effects of "AutoSense on" and who chose the "AutoSense off" setting. Therefore, it can be concluded that AutoSense may not be beneficial in all situations. In the study conducted by Wolfe et al. (2017), the majority of children with hearing aids also expressed a preference for "AutoSense on", although this was an older version of AutoSense. Our findings align with those of the aforementioned study.

## 4.3 General Discussion

In the current study, we evaluated the classification accuracy and transient behavior of AutoSense in different virtual acoustic environments. Moreover, we evaluated the effect of the directional microphone activated by AutoSense on the spatial awareness of CI children and on the speech intelligibility from the front in a more realistic virtual classroom situation incorporating diffuse classroom babble noise and reverberation. The new test paradigm using a VR headset with a virtual classroom environment proved highly effective. The enhanced ecological validity of the test conditions enabled more precise prediction of the potential benefits and drawbacks of AutoSense in real-world settings. Notably, the VR environment proved to be highly engaging for the participating children. All participants successfully completed the test, demonstrated concentration during the task, and reported enjoyment. Finally, real-world functional performance with and without AutoSense was compared in CI children. This comprehensive approach enables the formulation of evidence-based recommendations for clinical practice regarding the utilization of AutoSense in CI children.

As anticipated, AutoSense enhanced speech intelligibility from the front and influenced spatial awareness, albeit to a limited extent for the majority of directions. The sole exception was the 180° direction, where spatial awareness was significantly impaired. With regard to the remaining angles (0°, ±90° and ±120°), participants were still able to detect sounds and differentiate between words with "AutoSense on". This indicates that with AutoSense enabled, children should be able to detect someone calling their name from these directions. If they can infer the position of the new talker from the sound and/or visual cues, then they should be able to orient themselves to the new talker to benefit from the directional microphone activated by AutoSense. As CI children often sit in the front of the classroom, it is possible that some of their classmates are seated directly behind them. Therefore, during interactive lessons they may occasionally miss something their classmates say, but AutoSense will provide a benefit in most cases when the CI children turn their head towards the active speaker. CI children may need to be guided in their behavior to turn their head towards the person they want to hear. During group work their group members are at a close distance and unlikely to be located directly behind them, so in such situations AutoSense will probably also provide a benefit. During frontal instruction and individual work AutoSense might provide a benefit, but these are usually quieter situations and there will be no benefit when there is no noise. Given that group work and interactive lessons account for 34% of classroom time (Feilner et al., 2016), we expect AutoSense to provide a benefit in about 34% of classroom time, to be detrimental in a very small percentage of the time, and to make no difference during the rest of the classroom time. This is in accordance with the findings from Ricketts et al. (2017), who found that observers rated 33% of the classroom time to be optimal for a directional microphone setting. In the current study, no questions were included specifically about classroom situations in the questionnaires that participants had to fill in during the take-home period, so we do not know whether the benefit of AutoSense was noticeable. In future research, it would be advisable to include the teachers and teaching assistants in order to assess the perceived benefit of an automatic classifier at school.

In some other situations it may not be possible for CI children to turn their head, for example when they need to be focused on the ball during team sports or when they need to watch the road in front of them when navigating in traffic. In such situations, AutoSense will be detrimental if it activates the directional microphone. During the take-home period in the current study, one participant noticed a detrimental effect of AutoSense on the ability to understand teammates during sports, but a beneficial effect in other situations. Adding an additional manual program with an omnidirectional microphone mode for use during sports allowed this participant to benefit from AutoSense in other situations without the detrimental effect during sports. In traffic situations, it is still unclear whether it is more important for CI children to understand their parents or to hear approaching traffic. In the responses to the questionnaires in the current study, parents seemed to be more focused on speech intelligibility in traffic situations. This topic requires further research.

 It is recommended that clinicians provide CI children and their families with the opportunity to trial AutoSense Sky OS, as it may take time for users to adjust before the benefits in speech intelligibility become apparent. It is also important for clinicians to inform families about the adverse effects on spatial awareness, particularly if children frequently encounter situations in which they are addressed from behind, such as during sports activities. The creation of an additional manual program with an omnidirectional microphone mode for use during such situations is a viable option.

# 5 CONCLUSION

In the present study we were able to quantify the frontal speech intelligibility advantage of an automatic classifier in pediatric CI-users at school, at the cost of a slightly reduced spatial awareness. The general preference for AutoSense Sky OS among CI children and their parents appears to be a matter of individual trade-offs between enhanced speech intelligibility in noisy conditions and reduced spatial awareness. For the majority of participants, the advantages outweighed the disadvantages. Evidence-based recommendations for clinical practice regarding the utilization of AutoSense Sky OS in CI children were formulated.

# 6 ACKNOWLEDGEMENTS

The authors gratefully acknowledge the participation of the CI children and their parents. The authors would also like to thank Allart Knoop for developing the dummy head used in the technical evaluation, Noud Keijsers for his help with programming the CI processors of the participants, and Maryam Hussain for her help with conducting the listening tests.

This study was funded in part by Advanced Bionics AG. The funding organization had no role in the design and conduct of the study; in the collection, analysis, and interpretation of the data; or in the decision to submit the article for publication; or in the preparation, review, or approval of the article.

# 7 DATA STATEMENT

The data that support the findings of Experiment 2 of this study are available on request.

The room impulse response recordings made in the real classroom, the first-order ambisonics babble noise recording and the 3D environment of the classroom are published (Hendrikse & Tjepkema, 2025).

# 8 DECLARATION OF INTERESTS

☐ The authors declare that they have no known competing financial interests or personal relationships that could have appeared to influence the work reported in this paper.

☒ The authors declare the following financial interests/personal relationships which may be considered as potential competing interests:

Jantien Vroegop reports financial support and equipment, drugs, or supplies were provided by Advanced Bionics AG. If there are other authors, they declare that they have no known competing financial interests or personal relationships that could have appeared to influence the work reported in this paper.

# 9 DECLARATION OF GENERATIVE AI AND AI-ASSISTED TECHNOLOGIES IN THE WRITING PROCESS

During the preparation of this work the author(s) used ChatGPT, Copilot, and DeepL in order to improve language and readability, and to shorten the abstract. After using this tool/service, the author(s) reviewed and edited the content as needed and take(s) full responsibility for the content of the publication.

# APPENDIX

## APPENDIX A: TECHNICAL EVALUATION FOR PUB AND STREET ENVIRONMENTS

### Acoustic Conditions

In addition to the virtual classroom, two speech in noise situations in a pub and a traffic situation were used for the technical evaluation. The traffic situation was included to check for possible erroneous use of the directional microphone, as this can be detrimental for the ability to hear approaching traffic.

### 1. *Pub: conversation with a child in noise*

The pub is based on the pub environment and room impulse response recordings described in (Grimm et al., 2021; Van De Par et al., 2022). For more details, see Hendrikse et al. (2022; 2023). The target signal is a dialog between an adult and a child (-27° and -53°). Background noise includes babble noise and snippets of conversation from nearby tables. In addition, music is playing. Further away, at the bar, the sound of glasses clinking and noises from the beer tap can be heard. Finally, at certain times, there is the sound of chairs moving on the wooden floor. The babble noise of approximately 16 people speaking simultaneously was created by using the room impulse response recordings and conversation snippets from a database. The noise levels were selected to achieve a signal-to-noise ratio of approximately +3 $dB_A$ between the dialogue (69.4 $dB_A$; 72.2 $dB_C$) and background noise (66.4 $dB_A$; 70.1 $dB_C$). The layout and dimensions of the real pub were used to implement first-order reflections on the walls, floor, ceiling, and nearby tabletops. The room impulse response recordings were used to determine the settings of a feedback-delay network for the late reverberation, with a T60 of 0.65 s.

### 2. *Pub: conversation between three persons in loud noise*

The target signal is a conversation between three adults (-27°, +44°, +77°). The background noise is similar to that in the other pub situation, but with different babble noise and noise levels. The babble noise of approximately 54 people speaking simultaneously was created by using room impulse response recordings and conversation snippets from a database. The noise levels were selected to achieve a signal-to-noise ratio of approximately 0 $dB_A$ between the conversation (69.6 $dB_A$; 73.9 $dB_C$) and background noise (69.4 $dB_A$; 72.5 $dB_C$).

### 3. *Street*

The street is based on the acoustic street environment published by Grimm (2022). For more details, see Hendrikse et al. (2022; 2023). The listener is positioned at a street intersection where various vehicles are passing by, including cars, pickup trucks, and a bus. In addition, a bicycle approaches and rings its bell. People are passing by from behind (footsteps and conversations), and a mother with a baby in a pram passes by (mother singing, baby babbling), and she bumps over an empty can of soda. Finally, a car makes an emergency stop, and a train passes by at a railway crossing in the distance.

### Results and Discussion

Figures A1-3 show the acoustic conditions and the mean difference in dB with the reference recording for each of the classifier categories of AutoSense.

In the pub situations (Figures A1-2), the "speech in noise" category was active. We expected the "speech in loud noise" category to become active, but this did not happen with the default settings of AutoSense. In previous recordings with Naída M90-SP hearing aids equipped with AutoSense OS 3.0 (same as AutoSense Sky OS 3.0), we therefore lowered the activation level for the "speech in loud noise" category in the fitting software to the lowest possible setting to see what would happen. With the activation level at the lowest setting, AutoSense did activate the "speech in loud noise" category, but only in the pub situation with the more stationary babble noise (more simultaneous talkers) and lowest SNR (see Figure A2). The "speech in loud noise" category was always active together with the "speech in noise" category, because setting a marker in the "speech in noise" category affected the settings of both categories. AutoSense should switch to "speech in loud noise" when the noise floor is above 67 dB SPL for 15 seconds in its default setting. However, although the average noise levels were above 70 dB$_C$ SPL in both pub situations, AutoSense did not switch with the default activation level. We think that the real-life babble noise may have been too dynamic to cross the SPL threshold for the required amount of time.

In the street situation (Figure A3), AutoSense switched between the "calm situation" and "comfort in noise" categories, and switched to "speech in noise" when two talking persons were passing by. In all other situations, it makes sense that speech is prioritized over other sounds, but in traffic, activating the directional microphone could be detrimental for hearing approaching traffic. Perhaps there should be a separate classifier category for traffic situations so that this does not happen.

In some analysis windows, especially in the street situation, there is no data plotted for the difference between the reference and marker categories. This occurred because the mean difference over all octave bands did not cross the noise threshold we set for any of the classifier categories. This could happen if there was a quiet part in the acoustic input to the hearing aids, or if the acoustic input had a limited bandwidth and differences were only present in a few octave bands. However, looking at the mean difference over all octave bands was the most robust method, so we did not change it.

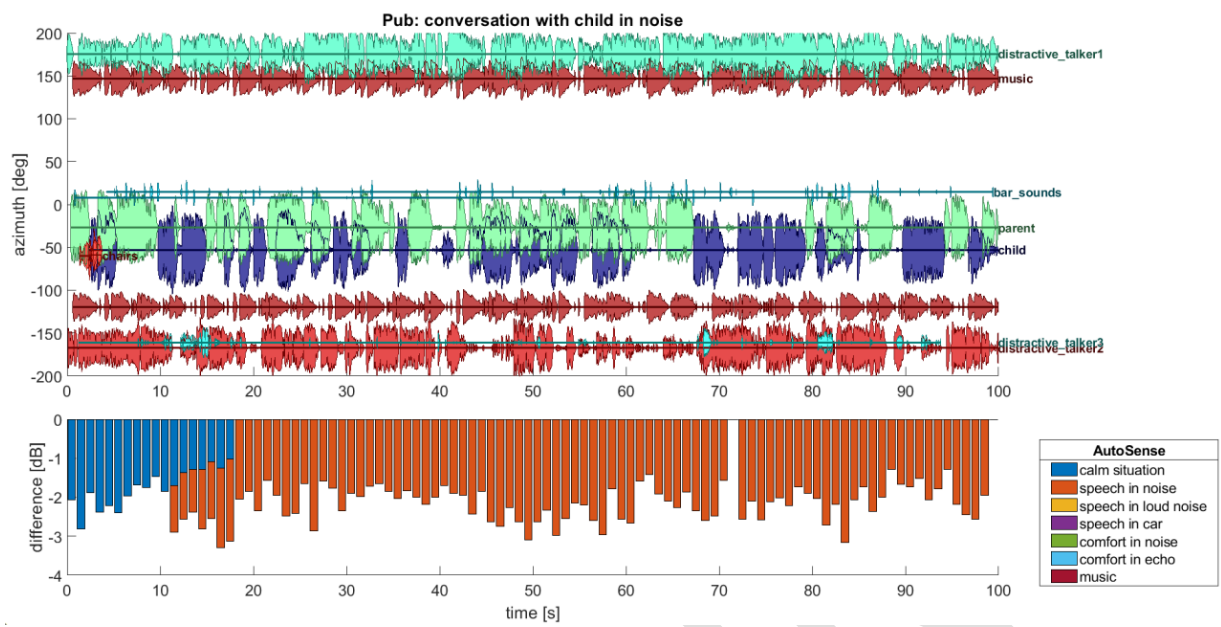

*Figure A1: Results of the technical analysis of the automatic classifier in the pub with the conversation with a child in noise. The top graph describes the acoustic conditions and shows the location in azimuth relative to the listener and the sound level (1 dB/degree) at the location of the listener. The bottom graph shows the difference in dB between the reference recording and the recordings with a marker in each of the classifier categories for Sky hearing aids with AutoSense Sky OS 3.0.*

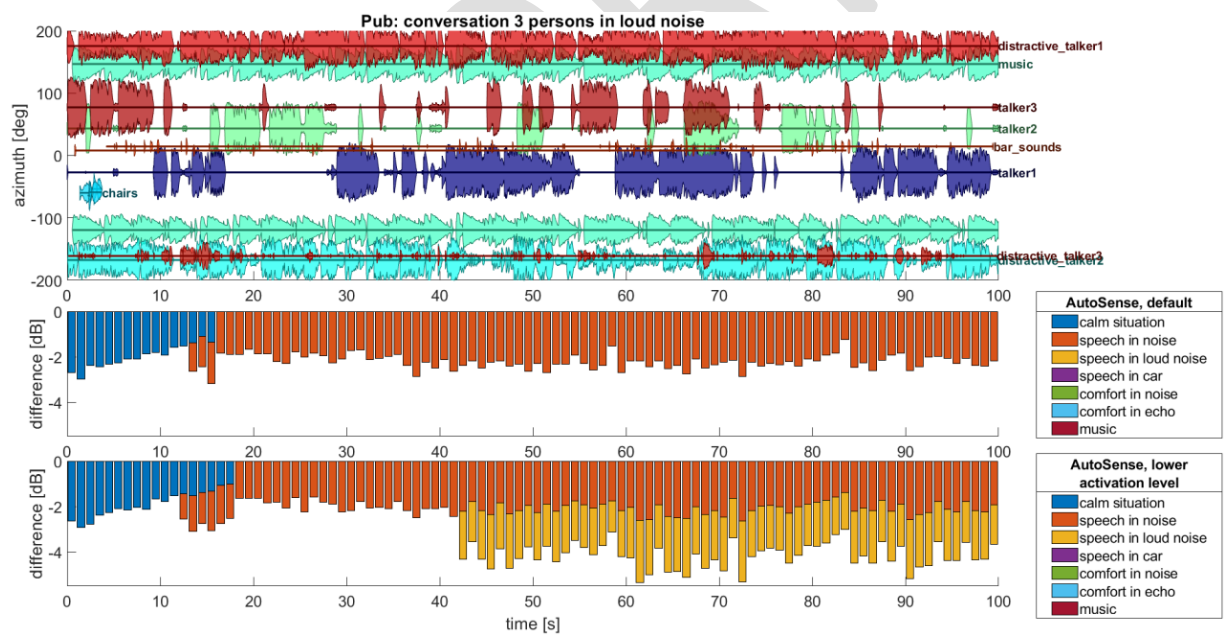

*Figure A2: Results of the technical analysis of the automatic classifier in the pub with a conversation between three persons. The top graph describes the acoustic conditions and shows the location in azimuth relative to the listener and the sound level (1 dB/degree) at the location of the listener. The middle graph shows the difference in dB between the reference recording and the recordings with a marker in each of the classifier categories for Sky hearing aids with AutoSense Sky OS 3.0. The bottom graph shows the difference in dB between the reference recording and the recordings with a marker in each of the classifier categories for Naída hearing aids with AutoSense OS 3.0 (same as AutoSense Sky OS 3.0) and the "speech in loud noise" activation level set to the lowest setting.*

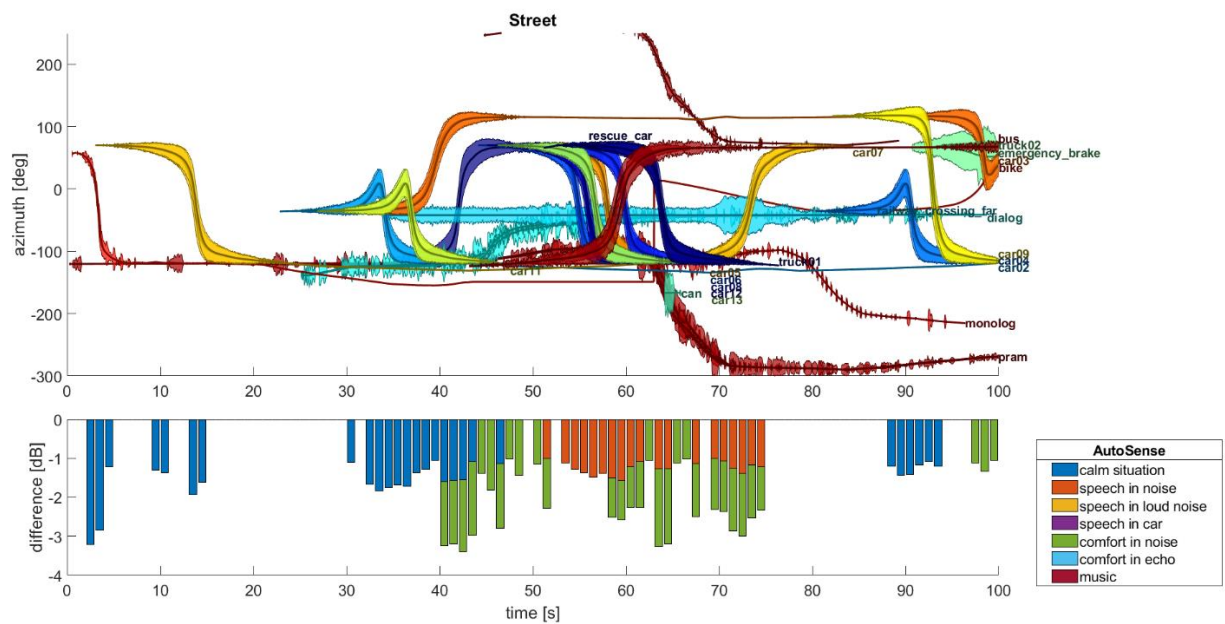

*Figure A3: Results of the technical analysis of the automatic classifier in the street. The top graph describes the acoustic conditions and shows the location in azimuth relative to the listener and the sound level (1 dB/degree) at the location of the listener. The bottom graph shows the difference in dB between the reference recording and the recordings with a marker in each of the classifier categories for Sky hearing aids with AutoSense Sky OS 3.0.*

## Appendix B: description of AutoSense (Sky) OS

AutoSense OS is the automatic classifier system used in Phonak hearing aids and Advanced Bionics cochlear implants. The pediatric (AutoSense Sky OS) and adult (AutoSense OS) versions of AutoSense are the same for version 3.0 and have the following categories:

- Calm situation
- Speech in noise
  - Speech in loud noise
  - Speech in car
- Comfort in noise
- Comfort in echo
- Music

"Speech in loud noise", "speech in car" and "music" are exclusive programs; while "calm situation", "speech in noise", "comfort in noise", and "comfort in echo" are non-exclusive blended programs. The "speech in loud noise" and "speech in car" categories are subcategories of "speech in noise", meaning that a change in the amplification settings of hearing aids of "speech in noise" also affects the amplification settings of the subcategories, but not vice versa. The "speech in loud noise" category activates a binaural beamformer (if applicable), with a stronger directionality than the "speech in noise" beamformer. AutoSense switches to "speech in loud noise" when both left and right hearing aid detect speech in noise at least 80% of the time, with a noise floor of at least 67 dB SPL for 15 seconds. This noise floor, called "activation level" in the fitting software, can be manually adjusted. The "comfort in echo" category has a reverberation suppression algorithm and can be active simultaneously with other categories.

For hearing aids, the amplification settings (gain-frequency curve) of the "calm situation" category are based on the hearing thresholds entered into the fitting software and one of the standard fitting rules. Amplification settings can be adjusted per category. By default, the other categories use the amplification settings of the "calm situation" category with some minor adjustments.

For cochlear implants, the amplification settings cannot be adjusted per category. The same map of electrode stimulation levels is used for all categories.

For both hearing aids and cochlear implants, the microphone directionality and the strength of some other noise reduction algorithms can be adjusted per category. By default, the microphone directionality is set to omnidirectional or "real ear sound" in the "calm situation" and "comfort in noise" categories, and directional in the "speech in noise" and "speech in loud noise" programs.